\documentclass[12pt,a4paper]{article}
\usepackage{graphicx,amssymb}
\usepackage{graphicx, array, epsfig, subfigure}
\usepackage{epstopdf}
\usepackage{subfigure}  
\newcommand{\be}{\begin{equation}}
\newcommand{\ee}{\end{equation}}
\newcommand{\bea}{\begin{eqnarray}}
\newcommand{\eea}{\end{eqnarray}}

\catcode`@=12

\begin{document}
\begin{center}
{\bf Effects of Violation of Equivalence Principle on UHE Neutrinos at IceCube 
in 4 Flavour Scenario}
\end{center}
\vspace{1cm}
\begin{center}
{ {\bf Madhurima Pandey} \footnote {email: madhurima.pandey@saha.ac.in}}
\end{center}
\vskip 0.5mm
\begin{center}
Astroparticle Physics and Cosmology Division  \\
Saha Institute of Nuclear Physics, HBNI  \\
1/AF Bidhannagar, Kolkata 700064, India  \\
\end{center}
\vskip 5mm
\begin{center}
{\small \bf Abstract}
\end{center}
{\small
If weak equivalence principle is violated then different
types of neutrinos would couple differently with gravity and that may 
produce a gravity induced oscillation for the neutrinos of different 
flavour. We explore here the possibility that very small violation 
of the principle of 
weak equivalence (VEP) can be probed by ultra high energy neutrinos from 
distant astrophysical sources. The very long baseline length and the 
ultra high energies of such neutrinos could be helpful to probe very small 
VEP. We consider a 4-flavour neutrino scenario (3 active + 1 sterile) with both 
mass-flavour and gravity induced oscillations and compare the 
detection signatures for these neutrinos (muon tracks and shower events) 
with and without gravity induced 
oscillations at a kilometer scale detector such as IceCube. We find that 
even very small VEP ($\sim 10^{-42}$) can considerably affect the detected 
muon yield produced by UHE neutrinos from distant Gamma Ray Bursts (GRBs)}. 

\newpage
\section{Introduction}
The oscillation of neutrinos \cite{bil} from one flavour to another are now established by several terrestrial 
experiments with neutrinos of natural origin such as solar and 
atmospheric neutrinos
and man-made neutrinos that include reactor \cite{reactorano,reactorano1,reactorano2} or accelerator neutrinos.
Due to the mass-flavour mixing of the neutrino eigenstates, 
a flavour eigenstate after traversing a distance can oscillate into
a different flavour due to the phase difference acquired by the 
mass eigenstates on its propagation. This phase depends on the mass squared 
differences and the baseline length. Thus discovery of the phenomenology of 
neutrino oscillation ensures that the neutrinos are massive. Being massive, 
The neutrinos should also therefore
undergo gravitational interactions. In the event that the weak interaction 
eigenstates of neutrinos are not the same as those of their gravity
eigenstates, neutrino oscillations can again be induced if different
neutrinos interact with gravity with different strengths, i.e. the 
gravitational constant $G$ is different for different types of neutrinos.
This situation may occur if the principle of weak equivalence is violated 
\cite{gas,gas1}.

General consequence of the weak equivalence principle is that there is
no difference between the gravitational mass and the inertial mass. This 
is to say that the force experienced by an object grounded on Earth is the 
same as the force experienced by the same object at the floor of a spaceship
which is moving with an acceleration same as that of the acceleration due to 
gravity in a no gravity environment. This can lead to the phenomenon of 
gravitational red shift $-$ under the influence of which the wavelength 
of a radiation suffers a widening (or the energy of a particle is shifted
towards a lower energy) while traversing through a gravitational field. 
The energies of the neutrinos too from a distant astrophysical object such as 
Gamma Ray Bursts (GRBs) would experience such a gravitational redshift on their 
travelling to the Earth. The shifted energy is given by 
$E' = (1 - \phi)E$ \cite{gas1,gas2}, where 
$\phi (= GM/R$) is the gravitational potential \cite{pot} through which the neutrino 
is propagating. If the equivalence principle is not violated then the energy
shifts for all the types of neutrinos are equal and this will not induce 
any phase difference between two
types of neutrinos during its propagation. But if the equivalence is 
violated then the energy shifts will be different for different types
of neutrinos (since the gravitational coupling $G_i (=G\alpha_i$, say) 
of the neutrino species
$i$ is different from $G_j (=G\alpha_j)$, the coupling for the species $j$). 
As a result, 
a pair of neutrino species ($i,j)$ will acquire a 
phase $\sim \Delta E L$ $(\Delta E = |E_i - E_j|$, $E_i$ and $E_j$ being 
the redshifted energies of the species $i$ and $j$ respectively) 
while traversing 
a distance $L$ (baseline length) from a distant
GRB, say, to Earth. Note that $E_i, E_j$ are the energy eigenstates 
in gravity basis. This would lead to a gravity induced oscillation
between neutrinos of different flavour 
with the oscillatory part given by 
$\sim |\Delta E L| = |\Delta f_{ij}| LE$ ($|\Delta f_{ij}| = |f_i - f_j|$)
with $f_i = (GM/R)\alpha_i$ $=\phi\alpha_i$.      

In general there are no specific signatures of violation of weak 
equivalence principle (VEP) in nature. But in case this is very  
weakly violated ($|\Delta f_{ij}|$ very small) then depending on the  
length of the baseline, neutrinos may probe such small VEP. For distant  
ultra high energy (UHE) neutrino sources such as GRBs, since the baseline 
length can be of the order of tens or hundreds of $\sim$ Mpc or more, 
gravity induced neutrino oscillations can be effective for very small 
violation of equivalence principle (such that the quantity 
$|\Delta f_{ij}| LE$ is not very small or very very large). 


In this work, we consider the UHE neutrinos from a GRB and estimate its
flux on reaching the Earth if they suffer both mass induced 
oscillations/suppressions and gravity induced oscillations. We then 
estimate the muon yields for these neutrinos at a 
kilometer square detector such as IceCube \cite{icecube} and compare our 
results 
with similar estimation when no oscillations are considered. 
For our estimation we consider a four flavour scenario where an extra sterile 
\cite{dunesterile}
neutrino is added with the three flavour families . 

The paper is organised as follows. In Section 2 we give the formalism for UHE 
neutrino fluxes from a single GRB (Subsection 2.1) as well as we discuss about 
the formalism of gravity and mass induced oscillations in the 4-flavour and 
3-flavour scenario (Subsection 2.2). In Section 3 we furnish calculational 
details and results. A discussion and summary is given in Section 4. 

\section{Formalism}
\subsection{Astrophysical neutrino fluxes from a single GRB}
Ultra high energy neutrinos may be produced via the highly energetic 
events in Gamma Ray bursts. Although the detailed mechanism of such  
high energy phenomena are yet to be understood, a well known model 
for GRB processes is the relativistically expanding fireball 
model. The fireball is generated and powered when a failed star or supernova 
leading to formation of a black hole accretes mass from its surroundings 
and the gravitationally infalling mass bounces back (like supernova explosion)
producing shock wave in the outward direction. This highly accelerated
outwardly mobile ``fireball" contains protons, $\gamma$ and carries with it 
enormous amount of energy. The proton - $\gamma$ interaction yields 
pions which decays to produce neutrinos through the process 
$\pi^+ \rightarrow \mu^+ + \nu_\mu$, $\mu^+ \rightarrow e^+ + \nu_e + 
{\bar\nu}_\mu$. For a three flavour case, neutrinos are then in the 
ratio $\nu_e : \nu_\mu : \nu_\tau  = 1:2:0$. If we consider a 
fourth sterile neutrino $\nu_s$ (as is considered in this work), 
then this ratio 
will be  $\nu_e : \nu_\mu : \nu_\tau : \nu_{\rm s} = 1:2:0:0$

The neutrino spectrum from a GRB can be parametrized as \cite{guetta,nayan1}
(with $E_{\nu}$, the neutrino energy and $N$ a normalization constant)
\bea
\displaystyle\frac {dN_{\nu}} {dE_{\nu}} = N \times {\rm min}
\left (1,\displaystyle\frac {E_{\nu}} {E_{\nu}^{\rm brk}} \right ) \displaystyle\frac {1} {E_{\nu}^2}\,\, .
\label{det1}
\eea
The spectral break energy $E_{\nu}^{\rm brk}$ in the above is related to 
the photon spectral break energy ($E_{\gamma, {\rm MeV}}^{\rm brk}$)
through the Lorentz boost factor ($\Gamma$) as
\bea
E_{\nu}^{\rm brk} \approx 10^6 \displaystyle\frac {\Gamma_{2.5}^2}
{E_{\gamma , \rm MeV}^{\rm brk}} \rm GeV\,\, ,
\label{det2}
\eea
where $\Gamma_{2.5}$ is the Lorentz boost factor $\Gamma$ normalized to 
$10^{2.5}$ ($\Gamma = \Gamma/10^{2.5}$) and 
\bea
N = \displaystyle\frac {E_{\rm GRB}}
{1 + \rm ln (E_{\nu \rm max} / E_{\nu}^ {\rm brk})}\,\, ,
\label{det3}
\eea
 where $\nu_{\rm max}$ is the upper cut-off energy of the neutrino spectrum.
For a GRB at a redshift $z$ is the observed neutrino energy reaching the 
Earth would be 
\bea
E_{\nu}^{\rm obs} &=& \displaystyle \frac {E_{\nu}}{(1 + z)}\,\, .
\label{det4}
\eea
and the upper cut-off $E_{\nu_{\rm max}}^{\rm obs}$ will then be 
\bea
E_{\nu_{\rm max}}^{\rm obs} &=& \displaystyle \frac {E_{\nu_{\rm max}}}{(1 + z)}
\,\, .
\label{det5} 
\eea
In the absence of neutrino oscillation, the flux reaching the Earth will 
then be written as, 
\bea
\displaystyle\frac {dN_{\nu}}{dE_{\nu}^{ \rm obs}} =
\displaystyle\frac
{dN_{\nu}} {dE_{\nu}} \displaystyle\frac {1} {4\pi L^2(z)} (1 + z)\,\, ,
\label{det6}
\eea 
where the baseline length $L(z)$ for a GRB at a redshift $Z$ is expressed as
\bea
r(z) = \displaystyle\frac {c} {H_0} \int_{0}^{z} \displaystyle\frac {dz'}
{\sqrt{\Omega_{\Lambda} + \Omega_{m} (1 + z')^3}}\,\, .
\label{det7}
\eea
where $\Omega_{\Lambda} = 0.685$, $\Omega_m = 0.315$ are the cosmological
parameters representing dark energy density and dark matter density normalized 
to critical density of the Universe and the Hubble parameter
$H_0 = 67.4$ $\rm Km\,\, \rm sec^{-1}\,\, \rm Mpc^{-1}$ is adopted 
for the present calculations. The velocity of light is denoted by $c$
in the above equation.

Assuming no CP violation, the neutrino spectra in Eq. (\ref{det1})  
${\cal F} (E_\nu) = \displaystyle\frac {dN_\nu}
{dE_\nu^{\rm obs}} = \displaystyle\frac {dN_{\nu + \bar{\nu}}} 
{dE_\nu^{\rm obs}}$ and therefore, we have the flux for neutrinos only, 
to be 0.5${\cal F}(E_\nu)$.

Since the neutrinos are produced at the source in flavour ($e$, $\mu$,
$\tau$ and sterile $s$) ratio $1:2:0:0$, the flux for each neutrino flavour
at the source can be written as, 
\bea
\phi_{\nu_e}^s  =  \displaystyle \frac{1} {6} {\cal F}(E_\nu)\,\, ,
\phi_{\nu_\mu}^s  =  \displaystyle \frac{2} {6} {\cal F}(E_\nu)  =  2\phi_{\nu_e}^s\,\, ,
\phi_{\nu_\tau}^s = 0\,\, ,
\phi_{\nu_{\rm s}}^s = 0\,\, .
\label{flux}
\eea

In the 4-flavour framework considered here, the neutrinos experience four 
flavour oscillations 
upon reaching the terrestrial detector from the astronomical extragalactic 
sources. The flux of neutrino flavours on reaching the Earth can be expressed 
as 
\bea
{F_{\nu_e}} &=& {P_{ee}}{\phi^s_{\nu_e}}
+ {P_{\mu e}}{\phi^s_{\nu_\mu}}\,\,  ,\nonumber\\
{F_{\nu_\mu}} &=& {P_{\mu \mu}}{\phi^s_{\nu_\mu}}
+ {P_{e \mu}}{\phi^s_{\nu_e}}\,\, , \nonumber\\
{F_{\nu_\tau}} &=& {P_{e \tau}}{\phi^s_{\nu_e}}
+ {P_{\mu \tau}}{\phi^s_{\nu_\mu}}\,\, ,\nonumber   \\
{F_{\nu_s}} &=& {P_{e s}}{\phi^s_{\nu_e}}
+ {P_{\mu s}}{\phi^s_{\nu_\mu}}\,\, ,
\label{flux5}
\eea
where $P_{\alpha \beta} (\alpha, \beta = e, \mu, \tau, s)$ is the oscillation 
probability between the flavours $\alpha$ and $\beta$ and $F_{\nu_\alpha}$ is 
the flux for the neutrinos $\nu_\alpha 
(\alpha = e,\mu, \tau,s)$ on reaching the Earth for the four flavour case.

\subsection{Contribution of VEP to the neutrino oscillation probability}

In case of a nonvanishing rest mass of neutrinos the weak and mass 
eigenstates are not necessarily identical, a fact wellknown in the quark sector 
where both types of state are connected by the Cabibbo-Kobayashi-Maskawa (CKM) 
matrix. This non zero mass nature of the neutrino allows for the 
phenomenon of neutrino oscillations, first given by Pontecorvo 
\cite{pont,pont1}, and it can be 
described by quantum mechanics. They are observable as long as the 
neutrino wave packets from a coherent superposition of states. Such 
oscillations among the different neutrino flavours do not conserve individual 
lepton numbers but only total lepton number. So that neutrino oscillation can 
be expressed as a quantum mechanical phenomenon wherby a neutrino created with 
a specific lepton family number (``lepton flavour") can later be measured to 
have a different lepton family number.

The n flavour eigenstate $|\nu_\alpha \rangle$ 
(with $\langle \nu_\beta|\nu_\alpha \rangle = \delta_{\alpha \beta} $), 
where n is an arbitrary number of orthonormal eigenstates, are connected to 
the nth mass eigenstate 
(with $\langle \nu_i|\nu_j \rangle = \delta_{ij}$) via a unitary matrix 
$U$
\bea
| \nu_\alpha \rangle &=& \displaystyle\sum_{i} U_{\alpha i} |\nu_i \rangle\,\, ,
\label{form1}
\eea
with
\bea
\displaystyle\sum_{i} U_{\alpha i} U_{\beta i}^{*} = \delta_{\alpha 
\beta},\,\, \displaystyle\sum_{\alpha} U_{\alpha i} U_{\alpha j}^{*} = 
\delta_{ij}\,\, .
\label{form2}
\eea
For the 4 (3 active +1 sterile) flavour scenario, the 
neutrino flavour eigenstates are related to the  mass eigenstates through a 
$4 \times 4$ unitary matrix given as 
\bea
\left (\begin{array}{c}
\nu_{e} \\
\nu_{\mu} \\
\nu_{\tau} \\
\nu_{s} \end{array} \right )
= \tilde{U}_{(4 \times 4)} \left (\begin{array}{c}
\nu_{1} \\
\nu_{1} \\
\nu_{3} \\
\nu_{4} \end{array} \right )\,\, 
= \left ( \begin{array}{cccc}
\tilde{U}_{e1} & \tilde{U}_{e2} & \tilde{U}_{e3} & \tilde{U}_{e4} \\
\tilde{U}_{\mu1} & \tilde{U}_{\mu2} & \tilde{U}_{\mu3} & \tilde{U}_{\mu4} \\
\tilde{U}_{\tau1} & \tilde{U}_{\tau2} & \tilde{U}_{\tau3} & \tilde{U}_{\tau4} \\
\tilde{U}_{s1} & \tilde{U}_{s2} & \tilde{U}_{s3} & \tilde{U}_{s4} \end{array}
 \right )
\left (\begin{array}{c}
\nu_{1} \\
\nu_{1} \\
\nu_{3} \\
\nu_{4} \end{array} \right )\,\,  ,
\label{form3}
\eea
where $\tilde{U}_{\alpha i}$ etc. indicate the elements of the 
Pontecorvo-Maki-Nakigawa-Sakata (PMNS) matrix \cite{pmns}. The PMNS matrix 
$\tilde{U}_{(4 \times 4)}$ can be generated by 
considering the successive rotations ($R$) in terms of mixing angles 
$\theta_{14},\,\theta_{24},\,\theta_{34},\,\theta_{13},\,\theta_{12},\,
\theta_{23}$ \cite{element,madhu1}
\bea
\tilde{U}_{(4 \times 4)} &=& R_{34}(\theta_{34})R_{24}(\theta_{24})R_{14}(\theta_{14})
R_{23}(\theta_{23})R_{13}(\theta_{13})R_{12}(\theta_{12})\,\,  ,
\label{form4}
\eea
Since we consider no CP violation in the neutrino sector, the CP phases are 
absent. The successive rotation terms ($R$) for 4-flavour case is 
written as
{\small
\bea
R_{34} (\theta_{34}) &=& \left (\begin{array}{cccc}
1 & 0 & 0 & 0 \\
0 & 1 & 0 & 0 \\
0 & 0 & c_{34} & s_{34} \\
0 & 0 & -s_{34} & c_{34} \end{array} \right )\,\, ,
R_{24} (\theta_{24})  =  \left (\begin{array}{cccc}
1 & 0 & 0 & 0 \\
0 & c_{24} & 0 & s_{24} \\
0 & 0 & 1 & 0 \\
0 & -s_{24} & 0 & c_{24} \end{array} \right )\, , \nonumber \\
R_{14} (\theta_{14}) &=&  \left (\begin{array}{cccc}
c_{14} & 0 & 0 & s_{14} \\
0 & 1 & 0 & 0 \\
0 & 0 & 1 & 0 \\
0 & 0 & 1 & 0 \\
-s_{14} & 0 & 0 & c_{14} \end{array} \right )\, ,
R_{12} (\theta_{12})  =  \left (\begin{array}{cccc}
c_{12} & s_{12} & 0 & 0 \\
-s_{12} & c_{12} & 0 & 0\\
0 & 0 & 1 & 0 \\
0 & 0 & 0 & 1\end{array} \right )\, , \nonumber \\
R_{13} (\theta_{13})  &=&  \left (\begin{array}{cccc}
c_{13} & 0 & s_{13} & 0\\
0 & 1 & 0 & 0\\
-s_{13} & 0 & c_{13} & 0 \\
0 & 0 & 0 & 1 \end{array} \right)\, ,
R_{23} (\theta_{23})  =  \left (\begin{array}{cccc}
1 & 0 & 0 & 0\\
0 & c_{23} & s_{23} & 0 \\
0 & -s_{23} & c_{23} & 0 \\
0 & 0 & 0 & 1 \end{array} \right)\,\, .
\label{form5}
\eea
}
With Eq. (\ref{form5}) $\tilde{U}_{(4 \times 4)}$ can be expressed as
{\small
\bea
\tilde{U}_{(4 \times 4)} &=& \left (\begin{array}{cccc}
c_{14} & 0 & 0 & s_{14} \\
-s_{14}s_{24} & c_{24} & 0 &c_{14}s_{24} \\
-c_{24}s_{14}s_{34} & -s_{24}s_{34} & c_{34} & c_{14}c_{24}s_{34} \\
-c_{24}s_{14}c_{34} & -s_{24}c_{34} & -s_{34} & c_{14}c_{24}c_{34}
\end{array} \right ) \times
\left (\begin{array}{cccc}
{{U}}_{e1} & {{U}}_{e2} & {{U}}_{e3} & 0 \\
{{U}}_{\mu1} & {{U}}_{\mu2} & {{U}}_{\mu3} & 0 \\
{{U}}_{\tau1} & {{U}}_{\tau2} & {{U}}_{\tau3} & 0 \\
0 & 0 & 0 & 1 \end{array} \right )
\eea
}
{\small
\bea
&=& \left (\begin{array}{cccc}
c_{14}{{U}}_{e1} & c_{14}{{U}}_{e2} & c_{14}{{U}}_{e3} & s_{14}  \\
& & & \\
-s_{14}s_{24}{{U}}_{e1}+c_{24}{{U}}_{\mu1} &
-s_{14}s_{24}{{U}}_{e2}+c_{24}{{U}}_{\mu2} &
-s_{14}s_{24}{{U}}_{e3}+c_{24}{{U}}_{\mu3} & c_{14}s_{24}  \\
&&& \\
\begin{array}{c}
-c_{24}s_{14}s_{34}{{U}}_{e1}\\
-s{24}s{34}{{U}}_{\mu1}\\
+c_{34}{{U}}_{\tau1} \end{array} &
\begin{array}{c}
-c_{24}s_{14}s_{34}{{U}}_{e2}\\
-s{24}s{34}{{U}}_{\mu2}\\
+c_{34}{{U}}_{\tau2} \end{array}  &
\begin{array}{c}
-c_{24}s_{14}s_{34}{{U}}_{e3}\\
-s{24}s{34}{{U}}_{\mu3}\\
+c_{34}{{U}}_{\tau3} \end{array}  &
c_{14}c_{24}s_{34}    \\
&&& \\
\begin{array}{c}
-c_{24}c_{34}s_{14}{{U}}_{e1}\\
-s_{24}c_{34}{{U}}_{\mu1}\\
-s_{34}{{U}}_{\tau1} \end{array}  &
\begin{array}{c}
-c_{24}c_{34}s_{14}{{U}}_{e2}\\
-s_{24}c_{34}{{U}}_{\mu2}\\
-s_{34}{{U}}_{\tau2} \end{array}  &
\begin{array}{c}
-c_{24}c_{34}s_{14}{{U}}_{e3}\\
-s_{24}c_{34}{{U}}_{\mu3}\\
-s_{34}{{U}}_{\tau3} \end{array}  &
c_{!4}c_{24}c_{34}  \end{array} \right )\,\,  ,
\label{form6}
\eea
}
where $U_{\alpha i}$ are the PMNS matrix elements for 3 flavour mixing matrix 
$U$, which 
can be expressed as \cite{mat,mat1,amit}
\bea
U &=& \left (\begin{array}{ccc}
c_{12}c_{13} & s_{12}s_{13} & s_{13} \\
-s_{12}c_{23}-c_{12}s_{23}s_{13} & c_{12}c_{23}-s_{12}s_{23}s_{13} &
s_{23}c_{13} \\
s_{12}s_{23}-c_{12}c_{23}s_{13} & -c_{12}s_{23}-s_{12}c_{23}s_{13} &
c_{23}c_{13}  \end{array} \right )\,\,  .
\label{form7}
\eea
In Eqs. (\ref{form5}-\ref{form7}) , $c_{ij} = \cos\theta_{ij}$ and 
$s_{ij} = \sin\theta_{ij}$, where $\theta_{ij}$ is 
the mixing angle between ith and jth neutrinos having mass eigenstates 
$|\nu_i \rangle$ and $|\nu_j \rangle$.

The time evolution equation of neutrino mass eigen state (for 4 neutrino case) 
is given by 
\bea
i \displaystyle\frac {d} {dt}\left (\begin{array}{c}
\nu_{1} \\
\nu_{2} \\
\nu_{3} \\
\nu_{4} \end{array} \right )
= H_m \left (\begin{array}{c}
\nu_{1} \\
\nu_{1} \\
\nu_{3} \\
\nu_{4} \end{array} \right )\,\,  
= \left ( \begin{array}{cccc}
E_1 & 0 & 0 & 0 \\
0 & E_2 & 0 & 0 \\
0 & 0 & E_3 & 0 \\
0 & 0 & 0 & E_4 \end{array}
 \right )
\left (\begin{array}{c}
\nu_{1} \\
\nu_{1} \\
\nu_{3} \\
\nu_{4} \end{array} \right )\,\,  .
\label{form8}
\eea
In the flavour basis with $|\nu_\alpha \rangle = \tilde{U}_{(4 \times 4)} 
|\nu_i \rangle$ the evolution equation takes the form  
\bea
i \displaystyle\frac {d} {dt}\left (\begin{array}{c}
\nu_{e} \\
\nu_{\mu} \\
\nu_{\tau} \\
\nu_{s} \end{array} \right )
&=& H' \left (\begin{array}{c}
\nu_{e} \\
\nu_{\mu} \\
\nu_{\tau} \\
\nu_{s} \end{array} \right )\,\, ,
\label{form9}
\eea
where 
\bea
H' &=& \tilde{U}_{(4 \times 4)} H_m \tilde{U}^{\dagger}_{(4 \times 4)} \,\, .
\label{form10}
\eea
For the case of relativistic meutrinos of momentum $p$, the energy eigen 
value for $|\nu_i \rangle$ can be expressed as 
\bea
E_i = \sqrt{p_i^{2} +m_i^{2}} \simeq p_i + \displaystyle\frac {m_i^{2}} {2 p_i} 
\simeq p + \displaystyle\frac {m_i^{2}} {2 E}\,\, ,
\label{form12}
\eea
where $p_i \simeq p,$ (i = 1,2,3,4) is assumed. Now by using Eq. (\ref{form12}) we can write $H_m$ as 
\bea
H_m &=& \left ( \begin{array}{cccc}
p & 0 & 0 & 0 \\
0 & p & 0 & 0 \\
0 & 0 & p & 0 \\
0 & 0 & 0 & p \end{array}
 \right ) + \displaystyle \frac {1} { 2 E}
\left (\begin{array}{cccc}
m_1^{2} & 0 & 0 & 0 \\
0 & m_2^{2} & 0 & 0 \\
0 & 0 & m_3^{2} & 0 \\
0 & 0 & 0 & m_4^{2} \end{array}
\right)\,\, .
\label{form13}
\eea
In the above Eq. (\ref{form13}), we can neglect the matrix ${\rm diag} 
(p,p,p,p)$ as it does not induce any phase differences between the neutrinos 
and hence does not contribute to the neutrino oscillations. Substract 
$m_1^{2}$ from all the diagonal elements of the 
matrix diag$(m_1^{2},m_2^{2},m_3^{2},m_4^{2})$, Eq. (\ref{form13}) takes 
the form 
\bea
H_m &=& \displaystyle\frac {1} {2 E} {\rm diag} (0,\Delta m_{21}^2,\Delta m_{31}^2,\Delta m_{41}^2)\,\, .
\label{form14}
\eea
As mentioned earlier, the oscillation of neutrinos can also be induced in 
case the weak equivalence principle is violated in nature. In such a scenario, 
the gravitational couplings to different types of neutrinos will be different. 
Therefore in 
this case the gravitational constant $G$ no more remains same for different 
types of neutrinos.   
If the neutrino eigenstates in gravity basis $|\nu_{G_i} \rangle$  
are not the same as the flavour eigenstates $|\nu_\alpha \rangle$ of 
neutrinos then this can lead to 
neutrino oscillations even though neutrinos are massless. In the present 
work however we consider $|\nu_\alpha \rangle \ne |\nu_i \rangle \ne 
|\nu_{G i} \rangle$ ($|\nu_{G i} \rangle$ being the gravity eigenstate 
for neutrino $i$) 
such that both the mass flavour oscillations and 
gravity induced oscillations are explored in a single framework.  

In general, no positive signatures have been found for the violation of the 
weak equivalence principle. In the event that the equivalence principle 
is violated by a very small account then this may be detected by studying 
the gravity induced oscillations of neutrino. The effect can be manifested 
for the neutrinos with the very long basline ($\sim {\rm Mpc}$). The UHE 
neutrinos 
from distant high energy extragalactic sources can well be a possibility 
to test the VEP. 

In the theory of general relativity the equivalence principle is the 
equivalence of gravitational and inertial mass. The gravitational 
``force" as experienced locally while standing on a 
massive body (such as the Earth) is the same as the pseudo force experienced 
by an observer in a noninertial (accelerated) frame of reference. 
Therefore equivalence principle is violated if the universality of the
gravitational constant $G$ is no more valid. A consequence of the equivalence 
principle is that an object with an energy $E$ in a gravitational field will 
suffer a shift in energy in the same way as would be observed in an 
accelerated frame of reference in a no gravity environment. If we assume 
a weak and static gravitational field then 
this can be shown that for 
such a field, the metric is diagonal with $g_{00} = (1 + 2\phi)$. 
with $\phi(= - \frac { G M} {R})$ is the gravitational potential and 
where $R$ is the distance over which the gravitation field is operational 
and $M$ is the mass of the source of the gravitational field. 
The energy from an object in this potential $\phi$ will be redshifted by an 
amount given by 
$E' = \sqrt{g_{00}} E = E(1 - \frac { G M} {R}) = E(1 + \phi)$ 
\footnote {The relation $E' = \sqrt{g_{00}} E$ can be realised by noting 
that the proper time in a curved manifold (presence of gravitation) is 
$d\tau = \sqrt{g_{\mu \nu} dx^{\mu}dx^{\nu}}$. Now the proper time is related 
to the coordinate time by $d\tau = \sqrt{g_{00}} dt$ (clock is at rest). If 
$N$ number of waves are emitted from a distant star with frequency 
$f_{\rm star}$ and proper time interval 
$\Delta {\tau_{\rm star}}$ and if the same are detected at Earth with 
frequency $f_{\rm Earth}$ with proper time interval $\Delta {\tau_{\rm Earth}}$ 
then $\displaystyle\frac {f_{\rm star}} {f_{\rm Earth}} = 
\displaystyle\frac {\Delta {\tau_{\rm star}}} {\Delta {\tau_{\rm Earth}}} = 
\displaystyle\frac {\sqrt{g_{00} (x_{\rm Earth})}} 
{\sqrt{g_{00} (x_{\rm star})}} = \sqrt{\left ( \displaystyle\frac 
{1+2 \phi_{\rm Earth}} {1+2 \phi_{\rm star}} \right )} = 1 + |\Delta {\phi|}$.}.

Let us consider the case where in a neutrino oscillation experiments, 
the neutrinos from a distant astrophysical object propagate through a 
gravitational field in addition to the vacuum. With respect to the vacuum, 
the neutrino 
energies are redshifted (due to the Doppler effect) by an amount 
$E \rightarrow E' = \sqrt{g_{00}} E$. But because of the universality  
of the gravitaional coupling, the equivalence principle indicates that 
for all the neutrino types the energy shifts should be the same and 
therefore it cann not lead to any neutrino oscillations. Only 
the non-universality of the coupling of gravity to the neutrino field, because 
of the consequence of the possible violation of equivalence principle, can 
contribute to the neutrino flavour oscillations.

In the presence of the gravitational field, the flavour eigenstates 
$|\nu_\alpha \rangle (\alpha = e, \mu, \tau, s)$ can be expressed as the 
superpositions of the gravitational eigenstate $|\nu_{G i} \rangle 
(i = 1,2,3,4)$ in terms of the mixing angle parameters 
$\theta_{ij}^{'} (i \ne j), i,j = 1,2,3,4$ in 
the 4-flavour framework.
\bea
|\nu_\alpha \rangle &=& \tilde{U}_{(4 \times 4)}^{'} |\nu_{G i} \rangle \,\, ,
\label{form15}
\eea
where $\tilde{U}_{(4 \times 4)}^{'}$ represents the flavour-gravity mixing matrix in 4-flavour scenario
{\small
\bea
\tilde{U}_{(4 \times 4)}^{'} &=& \left (\begin{array}{cccc}
c'_{14}{{U'}}_{e1} & c'_{14}{{U'}}_{e2} & c'_{14}{{U'}}_{e3} & s'_{14}  \\
& & & \\
-s'_{14}s_{24}{{U'}}_{e1}+c'_{24}{{U'}}_{\mu1} &
-s'_{14}s_{24}{{U'}}_{e2}+c'_{24}{{U'}}_{\mu2} &
-s'_{14}s_{24}{{U'}}_{e3}+c'_{24}{{U'}}_{\mu3} & c'_{14}s_{24}  \\
&&& \\
\begin{array}{c}
-c'_{24}s_{14}s'_{34}{{U'}}_{e1}\\
-s'{24}s'{34}{{U'}}_{\mu1}\\
+c'_{34}{{U'}}_{\tau1} \end{array} &
\begin{array}{c}
-c'_{24}s'_{14}s'_{34}{{U'}}_{e2}\\
-s'{24}s'{34}{{U'}}_{\mu2}\\
+c'_{34}{{U'}}_{\tau2} \end{array}  &
\begin{array}{c}
-c'_{24}s'_{14}s'_{34}{{U'}}_{e3}\\
-s'{24}s'{34}{{U'}}_{\mu3}\\
+c'_{34}{{U'}}_{\tau3} \end{array}  &
c'_{14}c'_{24}s'_{34}    \\
&&& \\
\begin{array}{c}
-c'_{24}c'_{34}s'_{14}{{U'}}_{e1}\\
-s'_{24}c'_{34}{{U'}}_{\mu1}\\
-s'_{34}{{U'}}_{\tau1} \end{array}  &
\begin{array}{c}
-c'_{24}c'_{34}s'_{14}{{U'}}_{e2}\\
-s'_{24}c'_{34}{{U'}}_{\mu2}\\
-s'_{34}{{U'}}_{\tau2} \end{array}  &
\begin{array}{c}
-c'_{24}c'_{34}s'_{14}{{U'}}_{e3}\\
-s'_{24}c'_{34}{{U'}}_{\mu3}\\
-s'_{34}{{U'}}_{\tau3} \end{array}  &
c'_{14}c'_{24}c'_{34}  \end{array} \right )\,\,  .
\label{form16}
\eea
}
The evolution equation for $|\nu_{G i} \rangle$ can be written as
\bea
i \displaystyle\frac {d} {dt} |\nu_{G i} \rangle &=& H_{G i} 
|\nu_{G} \rangle \,\, ,
\label{form17}
\eea
where $H_{G} = {\rm diag} (E_{G1}, E_{G2}, E_{G3}, E_{G4})$ for 4 neutrino 
framework. Therefore the evolution equation for the flavour eigenstate 
($|\nu_\alpha \rangle$) for the case of massless neutrinos is written as 
\bea
i \displaystyle\frac {d} {dt} |\nu_\alpha \rangle &=& 
\tilde{U}'_{(4 \times 4)} H_{G} \tilde{U}'^{\dagger}_{(4 \times 4)} 
|\nu_{\alpha} \rangle \,\, .
\label{form18}
\eea
As discussed earlier, in the absence of any violation of equivalence principle 
all the 
gravitational energy eigenvalues ($E_{G} = \sqrt{g_{00}} E = 
(1 - \frac{GM} {R}) E $) will not induce any phase difference to the neutrino 
eigenstate after the propagation. But if the equivalence principle is 
violated, the gravitational coupling $G$ is different for different types 
of neutrinos and in that case we have $H_{G} = {\rm diag} 
\left ( (1 - \phi \alpha_{1})E, (1 - \phi \alpha_{2})E, (1 - \phi \alpha_{3})E, (1 - \phi \alpha_{4})E) \right )$, where $\frac {G_{i} M} {r} = 
\frac {G M} {r} {\alpha_{i}}  = \phi \alpha_{i}$. Therefore this will induce 
the phase differences $\Delta E_{ij, G}$, where  

\be
\Delta {E_{ij, G}} = \frac{GM} {R} \Delta{\alpha_{ij}} E = \phi \Delta 
\alpha_{ij} E\,\, ,
\label{form19}
\ee
where $\Delta{\alpha_{ij}} = |\alpha_i - \alpha_j|$. In what follows we use 
$U'$ and $U$ to signify the mixing matrix $\tilde{U}'_{(4 \times 4)}$ and 
$\tilde{U}_{(4 \times 4)}$ respectively.
The effective Hamiltonian of the system, which includes the contribution of 
both mass and gravitational mixing terms, can then be written as
\bea
H'' &=& U H_m U^{\dagger} + U' H_G U'^{\dagger} \,\, .
\label{form20}
\eea
It may be noted that for the UHE neutrinos with energies TeV and above the 
matter effect (MSW effect) for the neutrinos passing through the matter 
will not have only significance on neutrino oscillations. Thus in case 
equivalence principle is violated, the UHE neutrino from distant astrophysical 
objects will suffer only vacuum and gravity induced oscillations/suppressions. 
In our formalism, we assume that the mixing angle between mass and the 
flavour states and the mixing angle between the flavour and the gravitational 
eigenstate are same, i.e. $\tilde{U}_{(4 \times 4)} = \tilde{U}'_{(4 \times 4)}
$. Then the Hamiltonian with this assumption takes the form
\bea
H'' &=& U (H_m + H_G) U^{\dagger} \,\, ,
\label{form21}
\eea 
where 
\bea 
H_G &=& {\rm diag} (0, \Delta{f}_{21} E, \Delta{f}_{31} E, \Delta{f}_{41} E)\,\, .
\label{form22}
\eea
In Eq. (\ref{form22}) $\Delta{f}_{ij} = \Delta{\alpha}_{ij} \phi, 
i,j =1,2,3,4$ and $\tilde{U}_{(4 \times 4)} = \tilde{U}'_{(4 \times 4)} U$. 
So finally $H''$ can be 
written as 
\bea
H'' &=& U \, {\rm diag} (0, \displaystyle\frac {\Delta{m}_{21}^{2}} {2E} + 
\Delta{f}_{21} E, \displaystyle\frac {\Delta{m}_{31}^{2}} {2E} +
\Delta{f}_{31} E, \displaystyle\frac {\Delta{m}_{41}^{2}} {2E} +
\Delta{f}_{41} E) U^{\dagger} \nonumber\\
&=& \displaystyle\frac {1} {2E} U \, {\rm diag} (0,\Delta{\mu}_{21}^{2}, 
\Delta{\mu}_{31}^{2}, \Delta{\mu}_{41}^{2}) U^{\dagger}\,\, ,
\label{form23}
\eea
where
\bea
\Delta{\mu}_{21}^{2} &=& \Delta{m}_{21}^{2} + 2\Delta{f}_{21} E^{2} \nonumber\\
\Delta{\mu}_{31}^{2} &=& \Delta{m}_{31}^{2} + 2\Delta{f}_{31} E^{2} \nonumber\\
\Delta{\mu}_{41}^{2} &=& \Delta{m}_{41}^{2} + 2\Delta{f}_{41} E^{2} \,\, .
\label{form24}
\eea

Generally, the oscillation probability from a neutrino $|\nu_\alpha \rangle$ of 
flavour $\alpha$ to a neutrino $|\nu_\beta \rangle$ of flavour $\beta$ can 
be expressed as \cite{prob,madhu} 
\bea
P_{\nu_\alpha \rightarrow \nu_\beta} &=& \delta_{\alpha\beta}
- 4\displaystyle\sum_{j>i} U_{\alpha i} U_{\beta i} U_{\alpha j} U_{\beta j}
\sin^2\left (\frac {\pi L} {\lambda_{ij}} \right )\,\, ,
\label{form25}
\eea
with $U_{\alpha i}$, etc. are the elements of 
PMNS mixing matrix. In Eq. (\ref{form23}) $L$ defines the baseline length 
for the neutrinos (which in the present case $\sim {\rm Mpc}$ for UHE 
neutrinos from distant GRBs) and $\lambda_{ij}$ is the oscillation length. 
In the presence of both mass and graviaty induced oscillations, 
$\lambda_{ij}$ is given by
\bea
\lambda_{ij} = \displaystyle\frac {4 \pi E} {\Delta{\mu}_{ij}^{2}} = 
\displaystyle\frac {4 \pi E} {\left ( \Delta{m}_{ij}^{2} + 2 \Delta{f}_{ij} 
E^{2} \right )} \,\, .
\label{form26}
\eea
Eq. (\ref{form23}) reduces to the form 
\bea
P_{\nu_\alpha \rightarrow \nu_\beta} &=& \delta_{\alpha\beta}
- 4\displaystyle\sum_{j>i} U_{\alpha i} U_{\beta i} U_{\alpha j} U_{\beta j}
S_{ij}^{2}\,\, .
\label{form27}
\eea
With $S_{ij}^{2} = \sin^2 \left [\left ( \displaystyle\frac {\Delta{m}_{ij}^{2}} {4 E} + \displaystyle\frac {\Delta{f}_{ij} E} {2} \right ) L \right ]$.
In the 4-flavour scenario, the mass and the gravity induced oscillation 
probabilities are therefore expressed as 
\bea
P_{ee}^4 &=& 1 - 4[|U_{e2}|^2|U_{e1}|^2 S_{21}^2 +
 (|U_{e3}|^2|U_{e1}|^2+
|U_{e3}|^2|U_{e2}|^2) S_{32}^2 \nonumber\\ 
&&+ (|U_{e4}|^2|U_{e1}|^2+
|U_{e4}|^2|U_{e2}|^2) S_{42}^2 + 
|U_{e4}|^2|U_{e3}|^2 S_{43}^2] \nonumber\\
P_{e\mu}^4 &=& 4[|U_{e2}||U_{\mu 2}||U_{e1}||U_{\mu 1}| S_{21}^2 + 
(|U_{e3}||U_{\mu 3}||U_{e2}||U_{\mu 2}| + 
|U_{e3}||U_{\mu 3}||U_{e1}||U_{\mu 1}|) S_{32}^2 + \nonumber\\
&&(|U_{e4}||U_{\mu 4}||U_{e2}||U_{\mu 2}| + |U_{e4}||U_{\mu 4}||U_{e1}||U_{\mu 1}|) S_{42}^2 + |U_{e4}||U_{\mu 4}||U_{e3}||U_{\mu 3}| S_{43}^2]\nonumber\\
P_{e\tau}^4 &=& 4[|U_{e2}||U_{\tau 2}||U_{e1}||U_{\tau 1}| S_{21}^2 + 
(|U_{e3}||U_{\tau 3}||U_{e2}||U_{\tau 2}| + 
|U_{e3}||U_{\tau 3}||U_{e1}||U_{\tau 1}|) S_{32}^2 +\nonumber\\ 
&&(|U_{e4}||U_{\tau 4}||U_{e2}||U_{\tau 2}| + |U_{e4}||U_{\tau 4}||U_{e1}||U_{\tau 1}|) S_{42}^2 + |U_{e4}||U_{\tau 4}||U_{e3}||U_{\tau 3}| S_{43}^2]\nonumber\\
P_{es}^4 &=& 4[|U_{e2}||U_{s 2}||U_{e1}||U_{s 1}| S_{21}^2 + 
(|U_{e3}||U_{s 3}||U_{e2}||U_{s 2}| + 
|U_{e3}||U_{s 3}||U_{e1}||U_{s 1}|) S_{32}^2 + \nonumber\\
&&(|U_{e4}||U_{s 4}||U_{e2}||U_{s 2}| + |U_{e4}||U_{s 4}||U_{e1}||U_{s 1}|) S_{42}^2 + |U_{e4}||U_{s 4}||U_{e3}||U_{s 3}| S_{43}^2]\nonumber\\
P_{\mu \mu}^4 &=& 1 - 4[|U_{\mu2}|^2|U_{\mu1}|^2 S_{21}^2 + (|U_{\mu3}|^2|U_{\mu1}|^2+
|U_{\mu3}|^2|U_{\mu2}|^2) S_{32}^2 \nonumber\\
&&+ (|U_{\mu4}|^2|U_{\mu1}|^2+
|U_{\mu4}|^2|U_{\mu2}|^2) S_{42}^2 + 
|U_{\mu4}|^2|U_{\mu3}|^2 S_{43}^2] \nonumber\\
P_{\mu\tau}^4 &=& 4[|U_{\mu2}||U_{\tau 2}||U_{\mu1}||U_{\tau 1}| S_{21}^2 + 
(|U_{\mu3}||U_{\tau 3}||U_{\mu2}||U_{\tau 2}| + 
|U_{\mu3}||U_{\tau 3}||U_{\mu1}||U_{\tau 1}|) S_{32}^2 + \nonumber\\
&&(|U_{\mu4}||U_{\tau 4}||U_{\mu2}||U_{\tau 2}| + |U_{\mu4}||U_{\tau 4}||U_{\mu1}||U_{\tau 1}|) S_{42}^2 + |U_{\mu4}||U_{\tau 4}||U_{\mu3}||U_{\tau 3}| S_{43}^2]\nonumber\\
P_{\mu s}^4 &=& 4[|U_{\mu2}||U_{s 2}||U_{\mu1}||U_{s 1}| S_{21}^2 + 
(|U_{\mu3}||U_{s 3}||U_{\mu2}||U_{s 2}| + 
|U_{\mu3}||U_{s 3}||U_{\mu1}||U_{s 1}|) S_{32}^2 + \nonumber\\
&&(|U_{\mu4}||U_{s 4}||U_{\mu2}||U_{s 2}| + |U_{\mu4}||U_{s 4}||U_{\mu1}||U_{s 1}|) S_{42}^2 + |U_{\mu4}||U_{s 4}||U_{\mu3}||U_{s 3}| S_{43}^2]\nonumber\\
P_{\tau \tau}^4 &=& 1 - 4[|U_{\tau2}|^2|U_{\tau1}|^2 S_{21}^2 + (|U_{\tau3}|^2|U_{\tau1}|^2+
|U_{\tau3}|^2|U_{\tau2}|^2) S_{32}^2 \nonumber\\
&&+ (|U_{\tau4}|^2|U_{\tau1}|^2+
|U_{\tau4}|^2|U_{\tau2}|^2) S_{42}^2 + 
|U_{\tau4}|^2|U_{\tau3}|^2 S_{43}^2] \nonumber\\
P_{\tau s}^4 &=& 4[|U_{\tau2}||U_{s 2}||U_{\tau1}||U_{s 1}| S_{21}^2 + 
(|U_{\tau3}||U_{s 3}||U_{\tau2}||U_{s 2}| + 
|U_{\tau3}||U_{s 3}||U_{\tau1}||U_{s 1}|) S_{32}^2 + \nonumber\\
&&(|U_{\tau4}||U_{s 4}||U_{\tau2}||U_{s 2}| + |U_{\tau4}||U_{s 4}||U_{\tau1}||U_{s 1}|) S_{42}^2 + |U_{\tau4}||U_{s 4}||U_{\tau3}||U_{s 3}| S_{43}^2]\nonumber\\
P_{ss}^4 &=& 1 - 4[|U_{s2}|^2|U_{s1}|^2 S_{21}^2 + (|U_{s3}|^2|U_{s1}|^2+
|U_{s3}|^2|U_{s2}|^2) S_{32}^2 \nonumber\\
&&+ (|U_{s4}|^2|U_{s1}|^2+
|U_{s4}|^2|U_{s2}|^2) S_{42}^2 +
|U_{s4}|^2|U_{s3}|^2 S_{43}^2] \,\, .
\label{form28}
\eea
Similar probability equations cn be easily written for 3-flavour case 
considering the 3-flavour PMNS matrix Eq. (\ref{form7}).

\subsection{Detection of secondary muons produced from neutrino-nucleon 
interactions of diffuse GRB sources} 

We consider here a kilometer square detector such as IceCube for the 
detection of ultra high energy neutrinos from the GRBs. Upward going muons 
observed by the Super - Kamiokande detector are produced by 
the interactions between high energy atmospheric neutrinos, such as UHE 
neutrinos from distant extragalactic sources namely GRBs and the rock around 
the detector. For the case of detection of UHE neutrinos at a Km$^2$ 
detector like IceCube, we are looking for these upward going muons, whose 
production depends on neutrino ($\nu_\mu$) charged current interactions 
($\nu_\mu + N \rightarrow \mu + X$). The most promising advantage of 
considering upward going muons is that it cannot be misidentified as muons 
created in cosmic ray showers in the atmosphere.

The rate of upward going muon event from single GRB neutrino depends on 
$\nu_\mu N$ cross-sections in two different ways namely

i) The interaction length, which is a function of the total cross-section, 
that leads the attenuation of the neutrino flux due to interactions in the 
Earth and

ii) The probability that the neutrino induced muon arriving at the detector 
with an energy larger than the threshold energy $E_{\mu}^{\rm min}$.

For the UHE neutrino flux, we can represent the attenuation of the neutrinos, 
reaching the terrestrial detector being unabsorbed by the Earth, by a shadow 
factor ($S_{\rm shadow} (E_\nu)$). For a particular GRB, the zenith angle 
$\theta_z$ is fixed. Thus the shadow factor for a single GRB is given by
\bea
S_{\rm shadow}(E_{\nu}) = {\rm exp}[-z(\theta_{z})/L_{\rm int}(E_{\nu})]\,\,  ,
\label{shadow}
\eea
where $z (\theta_{z})$ is the column depth for the incident zenith angle 
$\theta_z$ of the neutrinos
\bea
z(\theta_z) = \int \rho(r(\theta_z,l)) dl\,\, .
\label{coldepth}
\eea
In Eq. (\ref{coldepth}) $\rho(r(\theta_z,l))$ ($l$ is the path length of neutrino in the Earth) indicates the matter density 
profile inside the Earth. We have taken Preliminary Earth Model (PREM) 
\cite{prem} to 
express the matter density profile of the Earth in a more convenient way as we 
consider Earth as a spherically symmetric ball in our work (dense inner and 
outer core and a lower mantle having medium density).   

The interaction length ($L_{\rm int} (E_\nu)$) in Eq. (\ref{shadow}) 
can be expressed as 
\bea
L_{\rm int} (E_\nu) = \displaystyle\frac {1} {\sigma^{\rm tot}(E_{\nu}) N_A}\,\, ,
\label{lint}
\eea
where $\sigma^{\rm tot}$ corresponds to the total (charge current 
($\sigma_{\rm CC}$) + neutral current ($\sigma_{\rm NC}$)) cross-section and 
$N_A$ represents the Avogadro number $N_A$ $ (= 6.023 \times 10^{23}\rm mol^{-1} = 6.023 \times 10^{23}\rm  gm^{-1})$.

The probability $P_\mu (E_\nu; E_{\mu}^{\rm min})$ for a muon, produced due to 
charge current interactions of neutrinos, reaching the detector having energy 
above $E_{\mu}^{\rm min}$ is expressed as 
\bea
P_{\mu}(E_{\nu}; E_{\mu}^{\rm min}) = N_A \sigma^{\rm cc}(E_{\nu})\langle R(E_{\nu}; 
E_{\mu}^{\rm min})\rangle\,\,\, ,
\label{muprob}
\eea
where the average range of muon in rock ($\langle R(E_{\nu}; E_{\mu}^{\rm min})\rangle$) is given as \cite{muon} 
\bea
\langle R(E_{\nu}; E_{\mu}^{\rm min})\rangle &=& \displaystyle \frac {1}
{\sigma_{\rm CC}} \int_{0}^{(1-E_{\mu}^{\rm min}/E_{\nu})} dy R(E_{\nu}(1-y); 
E_{\mu}^{\rm min}) \times \displaystyle \frac {d\sigma_{\rm CC}(E_{\nu}, y)} 
{dy} \,\, .
\label{muonrange}
\eea
We can write $E_\mu$ in the place of $E_\nu (1 - y)$ in Eq. (\ref{muonrange}) 
as $y( = (E_\nu - E_\mu)/E_\nu)$, defines the fraction of energy lost by a 
neutrino having energy $E_\nu$ in the production of secondary muons having 
energy $E_\mu$ via charge current interactions. The muon range 
$R(E_{\mu}; E_{\mu}^{\rm min})$ in Eq. (\ref{muonrange}) can be wriiten as 
\bea
R(E_{\mu}; E_{\mu}^{\rm min}) = \int_{E_{\mu}^{\rm min}}^{E_{\mu}} \displaystyle\frac {dE_{\mu}} {\langle dE_{\mu}/dX \rangle} \simeq \displaystyle\frac{1} 
{\beta} \rm ln \left ( \displaystyle\frac {\alpha + \beta E_{\mu}}
{\alpha + \beta E_{\mu}^{\rm min}} \right )\,\, .
\label{range}
\eea
The energy loss rate of muon having energy is expressed as \cite{t.k} 
\bea
\left \langle \displaystyle\frac{dE_{\mu}} {dX} \right \rangle = - \alpha -
\beta E_{\mu}\,\, ,
\label{energyloss}
\eea
where the constant $\alpha$ stands for the energy losses and $\beta$ describes 
the catastrophic losses (namely bremsstrahlung, pair production and hadron 
production) respectively. Now these two constants we have considered in our 
work are 
for $E_{\mu}\,\, \leq \,\, 10^6$  \rm GeV \cite{a.dar}
\bea
\alpha = {2.033 + 0.077\,\,{\rm ln}[E_{\mu}(GeV)]} \times 10^3\,\,{\rm GeV}\,\,{\rm cm^2}\,\, {\rm gm^{-1}}\,\, ,\nonumber\\
\beta = {2.033 + 0.077\,\,{\rm ln}[E_{\mu}(GeV)]} \times 10^{-6}\,\,{\rm GeV}\,\,{\rm cm^2}\,\,{\rm gm^{-1}}\,\, ,
\label{constants}
\eea
and otherwise \cite{guetta}
\bea
\alpha &=& 2.033\times 10^{-3}\,\, {\rm GeV}\,\, {\rm cm^2}\,\, {\rm gm^{-1}}\, ,\nonumber\\
\xi &=& 3.9 \times 10^{-6}\,\,{\rm GeV}\,\, {\rm cm^2}\,\, {\rm gm^{-1}}\, .
\label{highconstants}
\eea
As has already been mentioned, the detection of $\nu_\mu$'s from a distant GRB 
sources 
can be estimated from the tracks of the secondary muons. The total number of 
secondary muon yields, which is a function of both 
$S_{\rm shadow} (E_\nu)$ and $P_{\mu} (E_\nu; E_{\mu}^{\rm min})$, can be 
detected in a detector such as IceCube of unit area is 
(\cite{rgandhi1}, \cite{t.k}, \cite{nayan1})
\bea
S  =  \int_{E_{\mu}^{\rm min}}^{E_{\nu_{\rm max}}} dE_{\nu} 
S_{\rm shadow}(E_{\nu}) P_{\mu}(E_{\nu}; E_{\mu}^{\rm min}) 
\displaystyle\frac {dN_{\nu}} {dE_{\nu}}\,\, .
\label{rate}
\eea
We replace $\displaystyle\frac {dN_{\nu}} {dE_{\nu}}$ in Eq. (\ref{rate}) 
by $F_{\nu_{\mu}}$, mentioned in Eqs. (3), (33). We also consider 
the production 
of muons via the decay channel $\nu_\tau \rightarrow \tau \rightarrow 
\bar{\nu_\mu} \mu \nu_\tau$ with probability 0.18. In such cases we can 
compute the muon events by solving Eqs. (34)-(43) numerically, where 
$\displaystyle\frac {dN_{\nu}} {dE_{\nu}}$ in Eq. (43) is equivalent to 
$F_{\nu_{\tau}}$ (Eqs. (3),(33)).

\section{Calculations and Results}
This work explores the possibility that the UHE neutrinos from distant
GRBs may be effective in probing even a very small violation of 
equivalence principle. As discussed earlier, if equivqlence principle 
is violated, the different coupling strengths of different types 
of neutrinos with gravity can in turn induce a gravity induced 
oscillation among neutrino flavours. 
In such a scenario we compute here the neutrino
induced muon yields for thesse neutrinos from single GRBs, in a square 
kilometer Cerenkov detector such as 
IceCube and compare  with the computed values for the same in case 
when the equivalence principle is not violated.  
We consider a 4-flavour (3 flavour + 1 sterile) neutrino 
framework for all our
estimations and then compare our results with similar calculations 
considering the usual 
three active flavour neutrino oscillation scenario. 

In the presence of gravity induced oscillations with the usual mass 
flavour oscillations, we can calculate the neutrino induced secondary muon 
yield at a Km$^2$ IceCube detector for 4-flavour UHE 
neutrinos by using Eqs. (\ref{det1}) - (\ref{form28}) in section 1.1 and 1.2 
and Eqs. (\ref{shadow}) - (\ref{rate}) in section 1.3. For this purpose, we 
consider upward going muons, which are produced due to charged current 
interactions of UHE neutrinos in the detector or in the rocks during their
passage to the detector. The 
threshold energy of the detector has been taken as 
$E^{\rm min}_{\mu}$ = 1 TeV.

In the present calculations, we have chosen the best fit values of the 
three active mixing angles given by $\theta_{12} = 33.48^{\circ}, 
\theta_{23} = 45^{\circ}, \theta_{13} = 8.5^{\circ}$. Different neutrino 
experimental groups such as MINOS \cite{minos}-\cite{Adamson:2016jku}, 
Daya Bay \cite{daya}-\cite{daya6}, Bugey \cite{bugey}, NOVA 
\cite{nova}-\cite{nova5} 
have put limits on the flavour mixing angles 
($\theta_{14}, \theta_{24}, \theta_{34}$) 
in 4-flavour scheme. The upper limits of the four flavour neutrino 
mixing angles have obtained by NOVA as $\theta_{24}\leq 20.8^0$
and $\theta_{34}\leq 31.2^0$ for $\Delta m_{41}^2=0.5$ eV$^2$. For the same 
value of $\Delta m_{41}^2$, MINOS has proposed the upper limits on 
$\theta_{34}$ and $\theta_{24}$ as $\theta_{24}\leq7.3^0$ and 
$\theta_{34}\leq26.6^0$. In addition to the above mentioned experimental 
groups the IceCube-Deepcore results \cite{Aartsen:2017bap} suggest 
$\theta_{24}\leq19.4^0$ and $\theta_{34}\leq22.8^0$ for $\Delta m_{41}^2=1$ 
eV$^2$. In the present work we adopt the ranges for $\theta_{24}, \theta_{34} $ and $\theta_{14}$ to be 
$2^0\leq\theta_{24}\leq20^0$, $2^0\leq \theta_{34}\leq20^0$ and 
$1^0\leq \theta_{14} \leq4^0$ respectively and the limits on $\theta_{14}$ 
is consistent with the combined results obtained from MINOS, Daya Bay 
and Bugey experiments. In our calcuation we consider 
$\theta_{14}, \theta_{24} $ and $\theta_{34}$ as $4^0, 6^0$ and $15^0$ 
respectively.

The main motivation of our work is to demonstrate the effects of the gravity 
induced 
oscillations, in case of a possible violation of weak equivalence principle.
Thus our approach is to consider both the usual mass-flavour 
oscillation/suppression and the gravity induced oscillation in the same 
framework. We have also considered a 4 flavour scenario where an extra 
sterile neutrino is included in the oscillation formalism. We then 
demonstrate the effect of both the 4-flavour scenario and 
VEP with no VEP case vis-a-vis the similar effects for the usual three 
flavour scenario. 
 
The calculations are performed  
for the representative values (of $\Delta f_{ij}$, the amount of VEP) 
$\Delta f_{21} = 10^{-43}, \Delta f_{32} = 10^{-42}, \Delta f_{41} = 10^{-43}, 
\Delta f_{43} = 10^{-42}$ and we have considered $\Delta m_{32}^2$ and 
$\Delta m_{21}^2$ as $\Delta m_{32}^2 = 7.0 \times 10^{-5} {\rm eV}^2$ 
(from solar neutrino oscillation experimental results) and 
$\Delta m_{32}^2 = 2.4 \times 10^{-3} {\rm eV}^2$
(from atmospheric neutrino oscillation results) respectively. 
The values of the mass 
square differences in the 4-flavour cases such as $\Delta m_{41}^2$ lie 
within the range $0.2\, {\rm eV}^2 \leq \Delta m_{41}^2 \leq 2\, {\rm eV}^2$ 
and we assume that $\Delta m_{32}^2 \simeq \Delta m_{31}^2 \simeq 
2.4 \times 10^{-3} {\rm eV}^2$ and  $\Delta m_{32}^2 \simeq \Delta m_{31}^2 
\simeq 1\,{\rm eV}^2$.

\begin{figure}[h!]
\centering
\subfigure[]{
\includegraphics[height=6.0 cm, width=7.5 cm,angle=0]{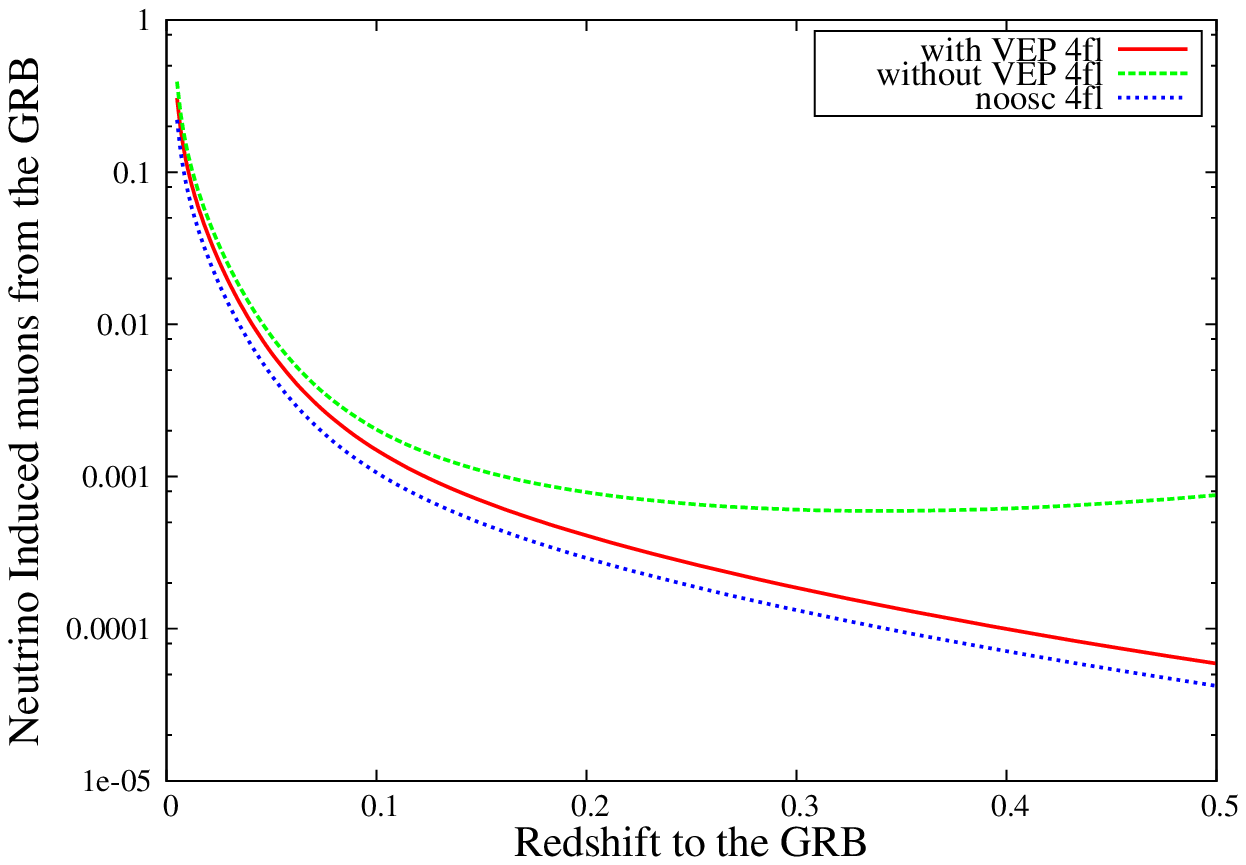}}
\subfigure []{
\includegraphics[height=6.0 cm, width=7.5 cm,angle=0]{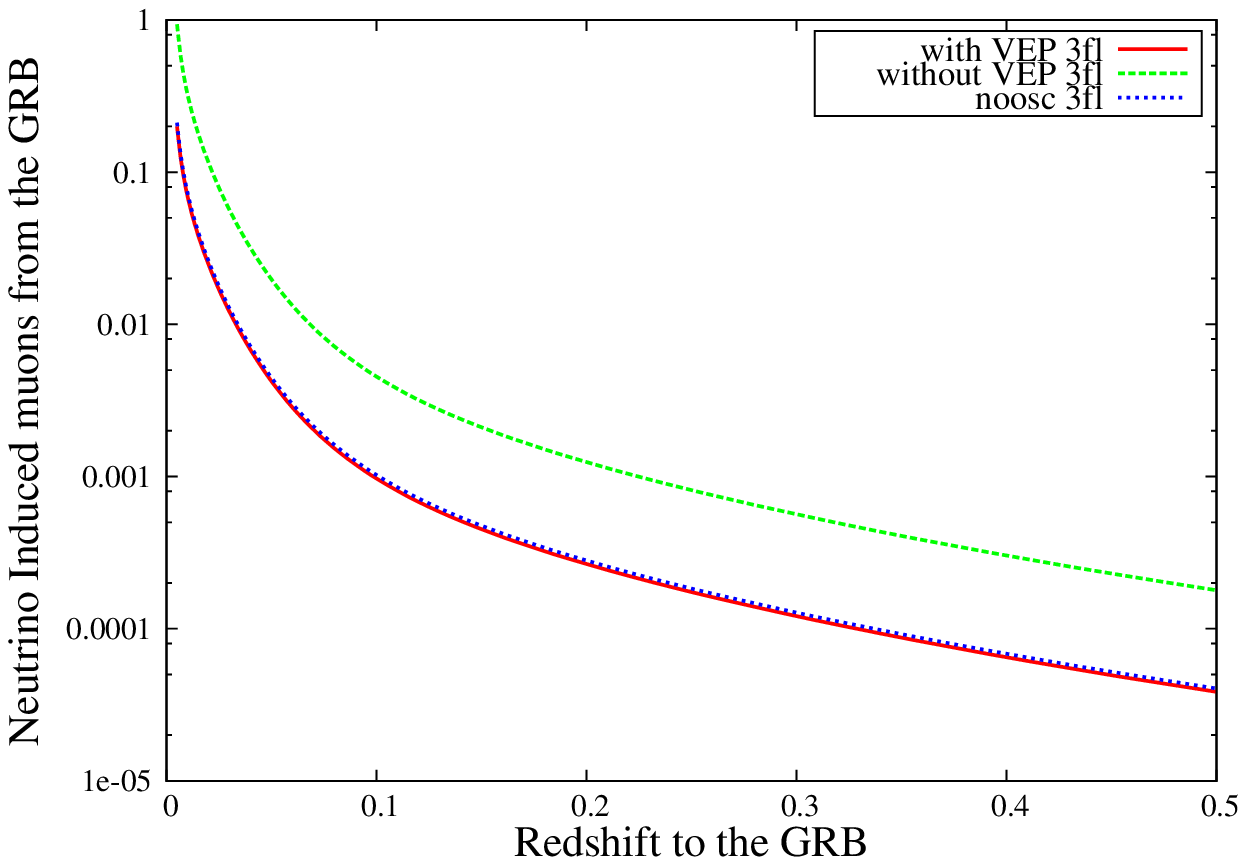}}
\caption{Variation of the muon yields from single GRB with different redshifts 
at a fixed zenith angle $\theta_z = 10^{0}$ for both 4-flavour (a) and 
3-flavour (b) cases. See text for details.}
\label{fig1}
\end{figure}
In Fig. 1, we furnish the variations of upward going muon yields 
at the IceCube like 
detector with different single GRBs at different redshift distances. 
The results are compared with the cases for only mass-flavour oscillation
(no gravity induced oscillations (i.e. no VEP)) and no oscillation 
(neither gravity induced nor mass induced). One can note from Fig. 1 that the 
muon yield with the gravity induced coupling differ with that of no VEP
scenario. Also the difference increses for GRBs at larger redshift distances
and at $z \sim 50$, the muon yields for the cases with no VEP 
differ by about 1 order of magnitude. We also show the effect of VEP 
oscillations while 3-flavour cases are considered in Fig 1(b). Similar feature 
is observed for the 3-flavour case too.  
\section{Summary and Discussions}

In this work, we explore the possibility that very small violation 
of equivalence principle can be probed via the gravity induced neutrino
oscillation. We demonstrate such a possibility by calculating the 
muon neutrino flux and consequently muon yields for such 
neutrinos at a kilometer square detector such as IceCube. We then compare 
our results for no oscillation case. We consider here a 4 neutrino scenario
and calculated the oscillation formalism with mass induced and gravity 
induced oscillations. We compare our results for gravity induced 
oscillations with a representative value of the VEP with those where
no VEP but only mass-flavour oscillation (suppression) is considered and also 
with no oscillation case for both 4-flavour and 3-flavour scenario.  
From comparisons of the muon yields we find that
UHE neutrinos from distant sources could be an effective way to probe 
very small violation of weak equivalence principle.    

{\bf Acknowledgments} : MP thanks the DST-INSPIRE fellowship grant by DST, Govt. of India.

\end{document}